\documentclass[12pt, times]{iopart}

\usepackage{physics}
\usepackage{graphicx}
\usepackage{afterpage}
\usepackage{array}
\usepackage[style=base]{caption}
\usepackage{subcaption}
\usepackage{geometry}
\usepackage{amsmath}

\newcommand{\RNum}[1]{\uppercase\expandafter{\romannumeral #1\relax}}

\begin{document}
\title[]{Highly Coherent Supercontinuum Generation in Circular Lattice Photonic Crystal Fibers Using Low-power Pulses}

\author{T. A. M. Ragib Shahriar$^{1}$, Ohidul Islam$^{1}$, Md Ishfak Tahmid$^{1, 2}$, Md. Zahangir Alam$^{2}$, and M. Shah Alam$^{2}$}

\address{$^1$Department of Electrical and Electronic Engineering, Shahjalal University of Science and Technology, Sylhet, Bangladesh \\
$^2$Department of Electrical and Electronic Engineering, Bangladesh University of Engineering and Technology, Dhaka, Bangladesh
}

\vspace{10pt}


\textbf{Abstract:} Two structures of circular lattice Photonic Crystal Fibers (PCFs) based on $\mathrm{Ge_{20}Sb_{15}Se_{65}}$ (GSS) material have been proposed for a highly coherent broadband supercontinuum generation (SCG) in the mid-infrared region. 
Numerical studies using the Finite Difference Eigenmode (FDE) solver on both structures show that the fundamental modes are well confined in the core while the confinement losses are very low. Also, the high nonlinear coefficient of 22.01 $\mathrm{W^{-1}m^{-1}}$ and 17.99202 $\mathrm{W^{-1}m^{-1}}$ for the two structures ensure that these structures can accomplish a high nonlinear activity. It has been found that broadband supercontinuums (SCs) spanning from 0.45 $\mathrm{\mu m}$ to 5.3 $\mathrm{\mu m}$  and 0.48 $\mathrm{\mu m}$ to 6.5 $\mathrm{\mu m}$ can be generated using a hyperbolic secant pulse of 0.5 kW. The proposed structures also show very good structural tolerance to optical properties that prevent any radical shift in SC spectra owing to potential fabrication mismatch. 

\vspace{0.5cm}

\section{Introduction}
Supercontinuum Generation (SCG) is an optical process of spectral broadening due to the interaction between dispersive properties and nonlinear properties of the medium ~\cite{dudley2006supercontinuum}. The SCG phenomena produce light sources with broad bandwidth, spatial coherence, and high brightness from ultrashort pulse which makes it useful for many applications. It has been widely researched recently due to its significant potential in various applications such as spectroscopy \cite{ke2009mid},  optical coherence tomography (OCT) ~\cite{marks2002study}, hyperspectral microscopy ~\cite{dupont2012ir}, and spectral tissue imaging \cite{petersen2018mid}. Supercontinuum (SC) has excellent applications in telecommunications systems also for Wavelength Division Multiplexing (WDM) by slicing into many wavelength channels ~\cite{boyraz200010,Ohara06}. It was first discovered by Alfano and Shapiro in bulk glass ~\cite{PhysRevLett.24.592,PhysRevLett.24.584} and since then it has been subject to numerous studies on nonlinear media.\\
In recent years many investigations have been made to find the best way to generate SC. Studies suggest that photonic crystal fibers (PCF) using materials of higher nonlinear coefficient are the best candidates as they can enable perfect control over nonlinear processes such as Stimulated Raman Scattering (SRS), soliton fission, Kerr effect, and Self Phase Modulation (SPM). Silica glass has been widely used but it gives low nonlinear effects ~\cite{begum2019near, FERHAT2018106, Jiang2015}. Chauhan \textit{et al.}~\cite{chauhan2019dispersion} generated supercontinuum using hexagonal photonic crystal design in which nonlinear coefficient was only 0.02627  $\mathrm{W^{-1}m^{-1}}$  and  Tarnowski \textit{et al.} ~\cite{tarnowski2016coherent} found nonlinear coefficient of 0.0077 $\mathrm{W^{-1}m^{-1}}$ using a hexagonal photonic crystal with  Germanium doped core. As chalcogenides have a high nonlinear coefficient, they can be considered as a good choice for SCG. Chalcogenides are compound component comprised of chalcogens from group \RNum{7} of the periodic table, including Sulfur (S), Selenium (Se), Tellurium (Te) with the addition of different components from group \RNum{15} like Arsenic (As) and Antimony (Sb), and group \RNum{14} elements like Germanium (Ge) and Silicon (Si) ~\cite{zakery2007optical}. The chalcogenide materials offer some alluring optical properties like high linear refractive index, high nonlinear refractive index, low photon energy, and enormous optical transparency extending from visible range to 20 $\mathrm{\mu m}$ ~\cite{jayasuriya2019mid}. Due to these properties of Chalcogenides, numerous computational and experimental works have been conducted to generate SCG with Photonic crystal fiber (PCF) using chalcogenide as background materials. Wang \textit{et al.} numerically showed that an SCG from 3866 nm to 5958 nm wavelength can be generated by pumping an ultrashort pulse at  4345 nm. In this study, they used a pulse of 4375 kW peak power in a 6 nm  PCF made of $As_2Se_3$  and they found the nonlinear coefficient larger than 1.0 $\mathrm{W^{-1}m^{-1}}$ ~\cite{wang2019flattened}. Saghaei \textit{et al.} numerically generated an ultra-wide mid-infrared SC in 50 mm step-index fiber (SIF) and photonic crystal fiber (PCF) using $As_{40}Se_{60}$ material. They determined the nonlinear coefficient around 0.5 $\mathrm{W^{-1}m^{-1}}$ for both fibers at pumping wavelength 4.25 $\mathrm{\mu m}$ ~\cite{saghaei2015ultra}. Gosh \textit{et al.} also reported mid-infrared supercontinuum generation spanning from 3.1 - 6.02 $\mathrm{\mu m}$  and 3.33 - 5.78 $\mathrm{\mu m}$ using femtosecond pumping at 4 $\mathrm{\mu m}$ and 4.53 $\mathrm{\mu m}$ using PCF of $As_{38}Se_{62}$ material ~\cite{ghosh2019chalcogenide}.\\
In this work, we numerically report on broadband SCG maximum spanning from  0.48 $\mathrm{\mu m}$ to 6.5 $\mathrm{\mu m}$ in a $Ge_{20}Sb_{15}Se_{65}$ (GSS) based  PCF. The dispersion engineering technique is incorporated to reduce the zero-dispersion wavelength (ZDW). First, we designed a single-mode PCF where air holes were placed in a circular fashion and analyzed dispersion, nonlinear coefficient, modal area, and other properties of PCF. In the second stage, we have used the designed PCF to investigate SC evolutions by solving the Nonlinear Schrödinger Equation (NLSE). A hyperbolic secant pulse of 0.5 kW peak power and 50 fs full-width half maximum (FWHM) was pumped as an ultrashort pulse in an anomalous dispersion regime.

\section{Design of the PCFs}
\begin{figure*}[ht!]
    \centering
      \begin{subfigure}[b]{0.35\textwidth}
        \includegraphics[width=\textwidth]{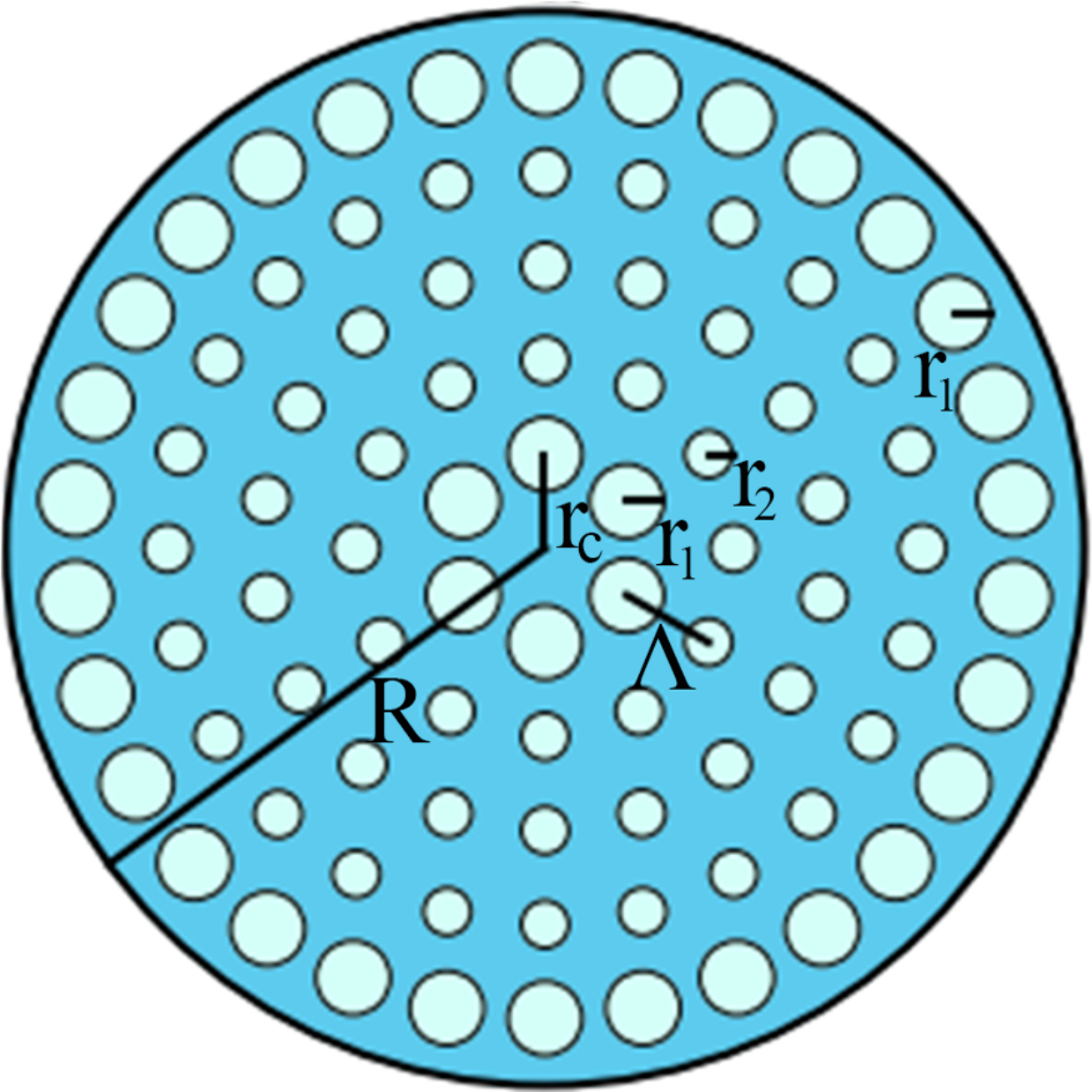}
        \caption{}
        \label{fig:design-1}
        \end{subfigure}
        \begin{subfigure}[b]{0.35\textwidth}
        \includegraphics[width=\textwidth]{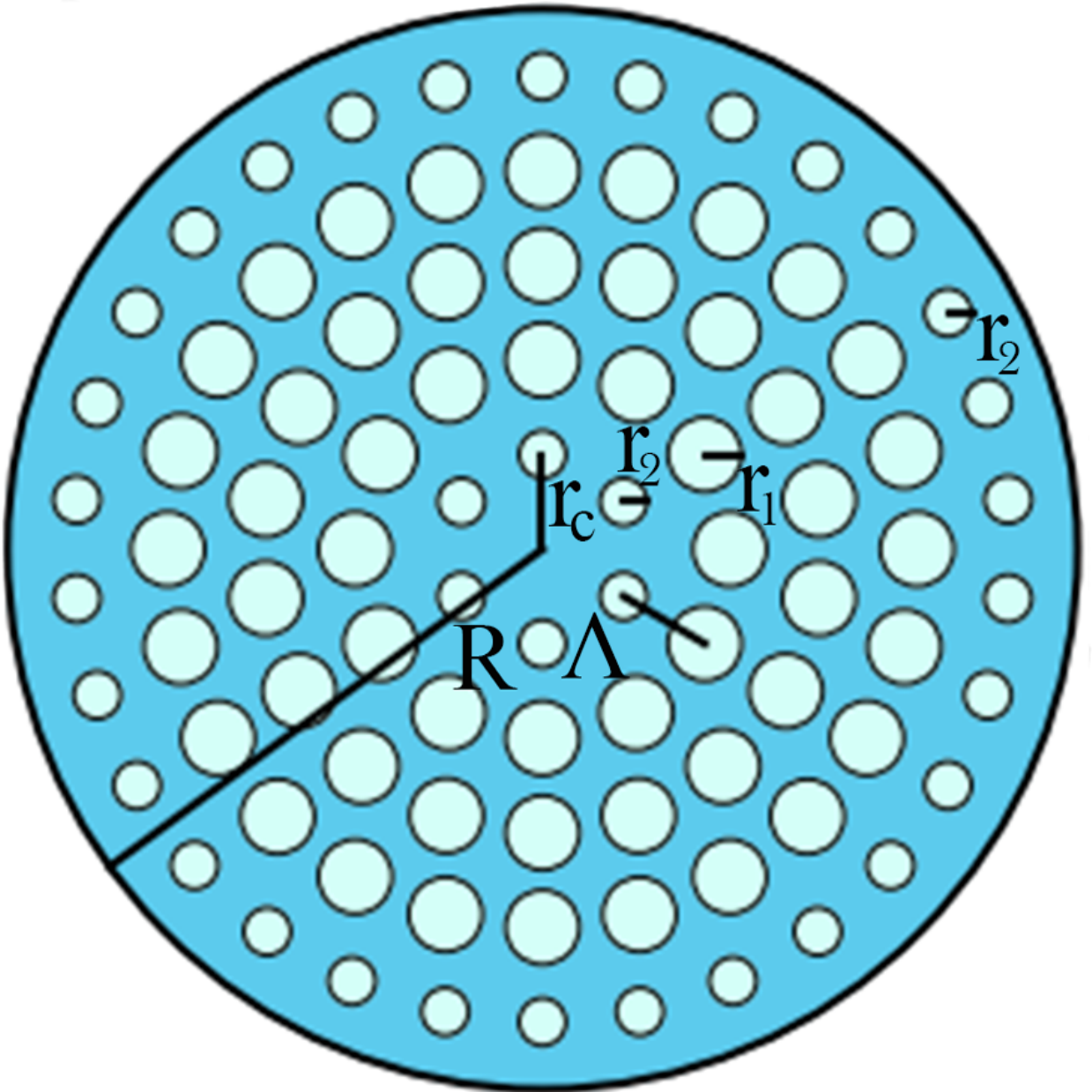}
        \caption{}
        \label{fig:design-2}
        \end{subfigure}
    
    \caption{Proposed designs of PCFs, (a) Structure-1 and (b) Structure-2.}
    \label{fig:design}
\end{figure*}

We have designed two solid core structures surrounded by five layers of air holes arranged in a circular fashion. The transverse cross sections are shown in figure \ref{fig:design}. As shown in the figure, the holes of the outermost and innermost layers have the same radius which is different than the radius of the holes of the middle three layers. We have swapped both radii to get dispersion at suitable points. We have optimized the parameters of the PCFs to obtain the desired dispersion and better confinement of the electric field in the core. The radius (R) of the structure is taken to be 8 $\mathrm{\mu m}$, the core radius is 1.4 $\mathrm{\mu m}$, and the pitch distance ($\Lambda$) from one hole layer to the next is 1.4 $\mathrm{\mu m}$. The radius of bigger and smaller air holes are depicted by $r_1$ and $r_2$ which are 0.5 $\mathrm{\mu m}$ and 0.4 $\mathrm{\mu m}$ respectively.

\noindent These design parameters are nearly close to the recently fabricated devices \cite{chaitanya2016ultra}. The GSS is chosen as background material due to its high nonlinearity. GSS is a dispersive material i.e. it’s refractive index varies with wavelength of the incident light. We used the Sellmeier equation to evaluate it's wavelength dependent refractive index as

\begin{align}
n(\lambda) = \sqrt{A_0+\frac{B_1\lambda^2}{\lambda^2-C_1}+\frac{B_2\lambda^2}{\lambda^2-C_2}+\frac{B_3\lambda^2}{\lambda^2-C_3}}.
\label{eq:sellmeier}
\end{align}

\begin{figure*}[ht!]
    \centering
      \begin{subfigure}[b]{0.35\textwidth}
        \includegraphics[width=\textwidth]{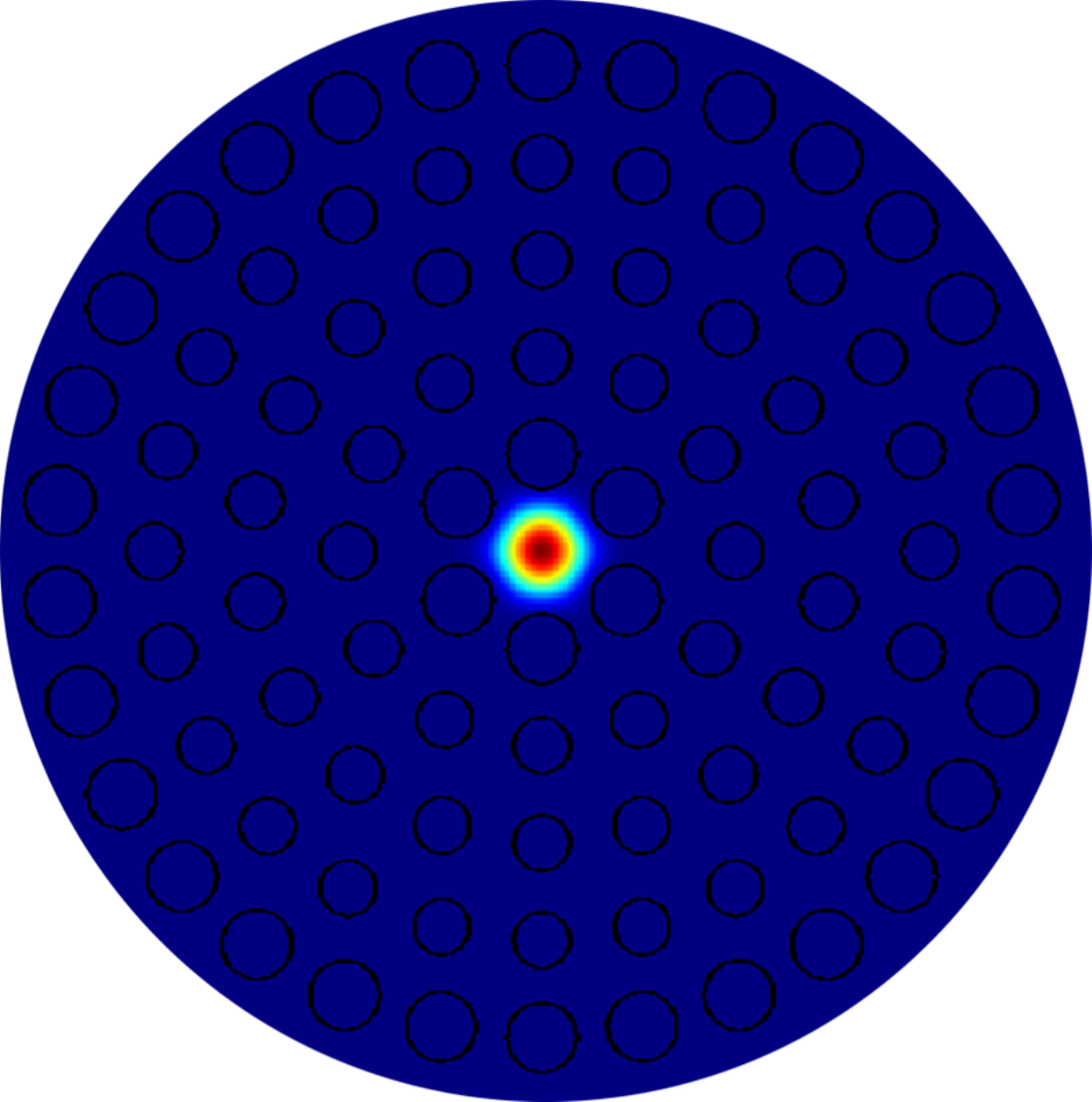}
        \caption{}
        \label{fig:mode-1}
        \end{subfigure}
        \begin{subfigure}[b]{0.35\textwidth}
        \includegraphics[width=\textwidth]{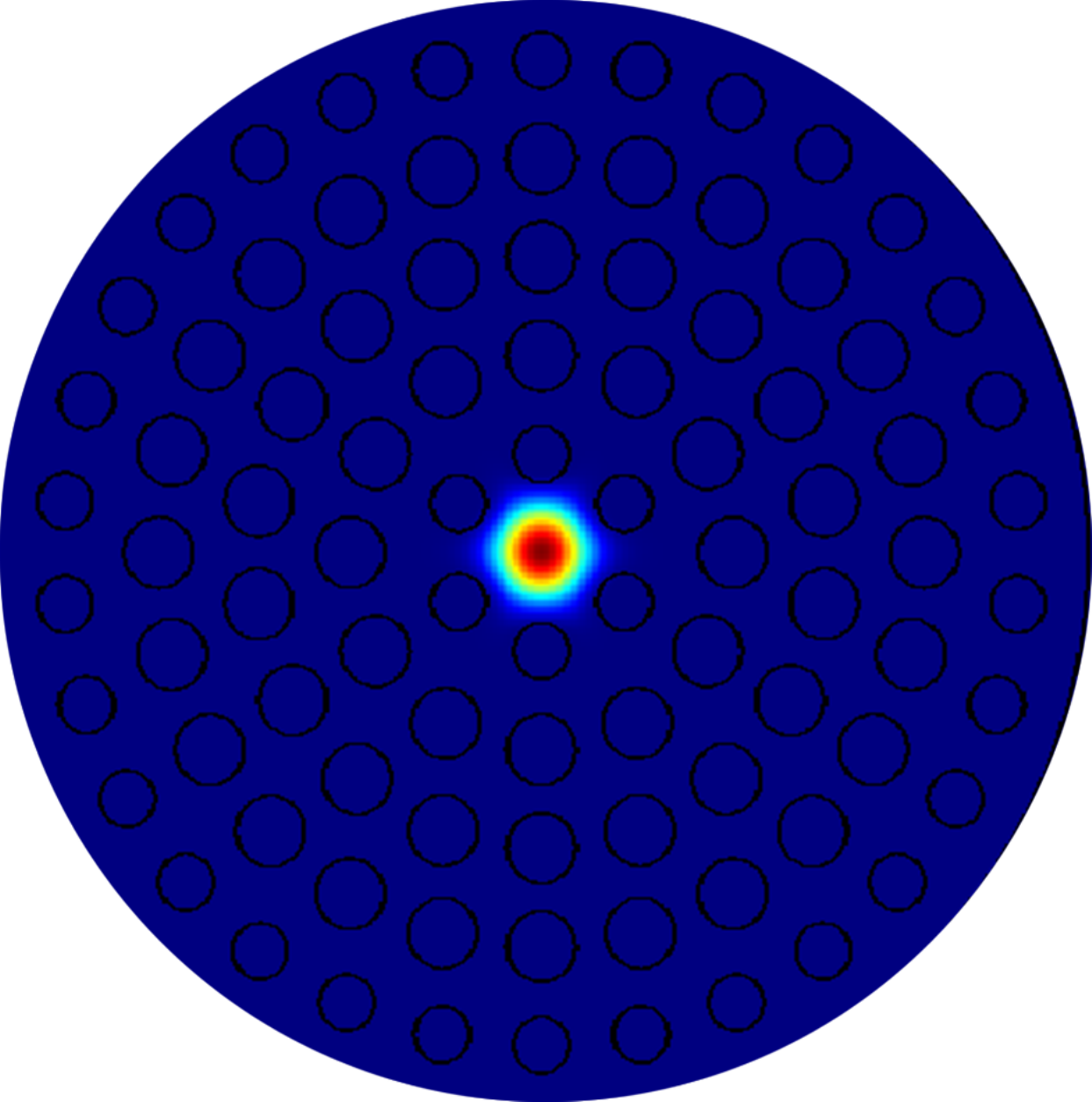}
        \caption{}
        \label{fig:mode-2}
        \end{subfigure}
    
    \caption{Electric field distribution of fundamental mode  at 1.2 $\mu$m wavelength, (a) Structure-1 and (b) Structure-2.}
    \label{fig:mode}
\end{figure*}

\noindent Here, $A_0 = 3.8667$, $B_1 = 0.1366$, $C_1= 0.0420$ $\mu$m, $B_2 = 2.2727$, $C_2 = 0.01898$ $\mu$m, $B_3 = 0.0138$, and $C_3 = 68.8303$ $\mu$m, and $\lambda$ is in $\mu$m \cite{zhanqiang2018mid,amiri2019design}. To compute the field distribution and frequency-dependent optical properties in the PCFs, we have used the Lumerical, which is a finite difference eigenvalue (FDE) solver, where wave equations in terms of transverse electromagnetic fields are employed \cite{Zhu:02}. In the numerical analysis, we have used the perfectly matched layer (PML) as the boundary condition, which completely absorbs light at the boundary. The space and time dependence of electric field (\textbf{E}) and magnetic field (\textbf{H}) in the PCFs are given by

\begin{align}
    \begin{aligned}
         \textbf{E}(z,t) &=  \textbf{E}_0\exp{i(-\omega t + \beta z)} \\
     \textbf{H}(z,t) &= \textbf{H}_0\exp{i(-\omega t + \beta z)}
    \end{aligned}
    \label{eq:E_B}
\end{align}

\noindent where $\omega$ is the angular frequency and $\beta$ is the propagation constant. The guided structure is uniform in the direction of propagation, which is \textit{z}-axis in this case.
The electric field distributions of the fundamental mode of the two structures at 1.2 $\mu$m pump wavelength are represented in figure \ref{fig:mode}. 

\section{Optical Properties}
\subsection{V-Parameter and Confinement Loss}
V-Parameter of the photonic crystal fiber can be found using \cite{Nielsen:03},

\begin{equation}
    V(\lambda) = \frac{2\pi }{\lambda}a_{eff} \sqrt{n^{2}_{co}(\lambda) - n^{2}_{FSM}(\lambda)}.
    \label{eq:V_para}
\end{equation}

\begin{figure*}[ht!]
    \centering
        \begin{subfigure}[b]{0.35\textwidth}
        \includegraphics[width=\textwidth]{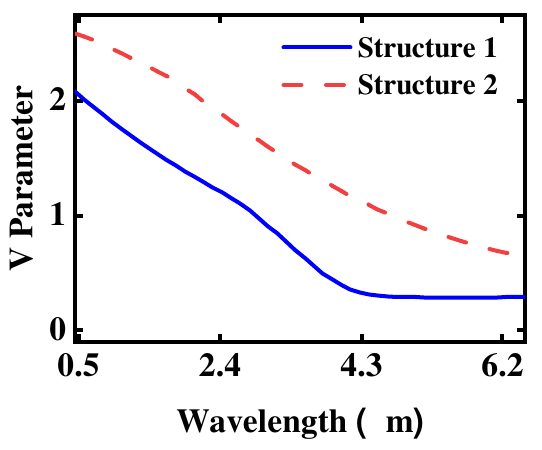}
        \caption{}
        \end{subfigure}
        \begin{subfigure}[b]{0.38\textwidth}
        \includegraphics[width=\textwidth]{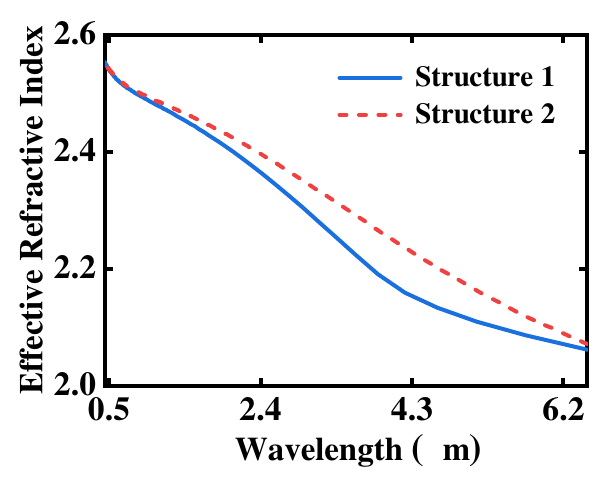}
        \caption{}
        \end{subfigure}
        
        \begin{subfigure}[b]{0.4\textwidth}
        \includegraphics[width=\textwidth]{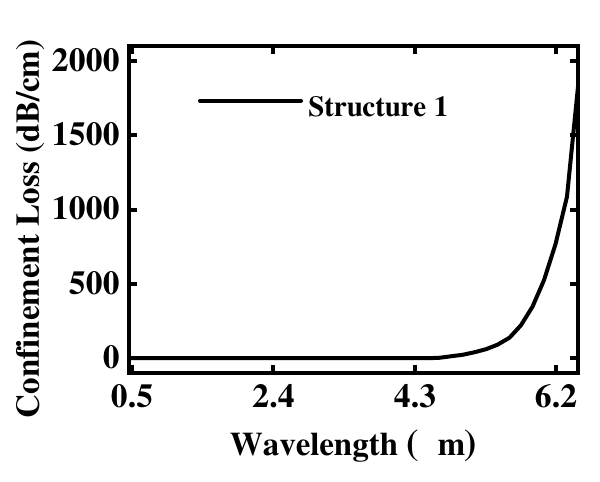}
        \caption{}
        \end{subfigure}
        \begin{subfigure}[b]{0.4\textwidth}
        \includegraphics[width=\textwidth]{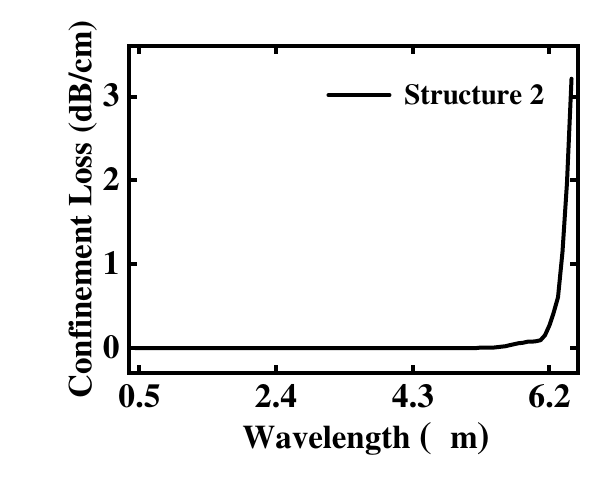}
        \caption{}
        
        \end{subfigure}
    \caption{(a) V-parameter of the PCFs. (b) The effective refractive index of both structures. Confinement loss of the structures, (c) Structure-1. (d) Structure-2.}
    \label{fig:v-param-conf-neff}
\end{figure*}

\noindent Here, $n_{co}$ is the effective refractive index of the fundamental mode, $n_{FSM}$ is the effective index of the fundamental space-filling mode which is considered as the effective index of the first mode that goes to the cladding in the infinite periodic cladding structure and $a_{eff}$ is the effective core radius of the fiber. Figure \ref{fig:v-param-conf-neff}(a) shows the V parameter for both structures. As equation (\ref{eq:V_para}) states that the V parameter will decrease with the increase of the operating wavelength, figure \ref{fig:v-param-conf-neff}(a) reveals the same. For single-mode optical fiber, the V parameter has to be less than 2.405 \cite{jurgensen1975dispersion}. At the pumping wavelength of 1.2 $\mu$m, the V parameters are found to be 1.41 and 2.32 for structure-1 and structure-2, respectively. So, both the PCFs are single-mode fibers that can support only the fundamental mode. The effective refractive index is defined as the ratio of the propagation constant of the fundamental mode $(\beta)$ to the propagation constant in vacuum ($k_0$) as ~\cite{senior2009optical},

\begin{align}
    n_{eff} = \frac{\beta}{k_0}
    \label{eq: neff}
\end{align}

\noindent Equation (\ref{eq: neff}) indicates that the vacuum wavelength is higher than the wavelength of the fundamental mode by the factor of $n_{eff}$. Change of the $n_{eff}$ with wavelength for both of the structures is shown in figure \ref{fig:v-param-conf-neff}(b). From the figure, we can see that the effective refractive index decreases as we move to a higher wavelength in both structures.

\noindent Confinement losses are the losses arising from the leaky nature of the modes and poorly designed core of photonic crystal fiber. To get an overview of how finely modes are confined in the core, it can be calculated using ~\cite{ALAM2021127322, Zahid_JNP},

\begin{equation}
    \alpha \mathrm{(dB/cm)} = 8.686 \times \mathrm{k_{0}Im(n_{eff})} \times 10^{4},
    \label{eq:confinement}
\end{equation}

\noindent where $k_{0} = 2\pi/ \lambda$ and Im$(n_{eff})$ is the imaginary part of the effective refractive index. The confinement losses for both structures are illustrated graphically in figure \ref{fig:v-param-conf-neff}(a) and figure \ref{fig:v-param-conf-neff}(b). It can be seen that the confinement loss is almost zero at the pumping wavelength. But, it increases in the longer wavelength region which indicates that pulse with the longer wavelength can not be confined well in the fiber core.
At an wavelength of 1.2 $\mu$m, the values of $\alpha$ for structure-1 and structure-2  are 6.14 $\times 10^{-7}$ $\mathrm{dB/cm}$ and $1.29\times 10^{-8}~\mbox{dB/cm}$, respectively. This low confinement loss indicates that modes are well confined in the core and very low leakage of electric flux occurs.

\subsection{Dispersion Coefficients}
The chromatic dispersion of a fiber is the sum of material dispersion and waveguide dispersion. By changing the geometrical parameters of the PCFs, desired dispersion can be engineered ~\cite{senior2009optical,kalantari2018ultra}.
\begin{figure*}[ht!]
    \centering
        \begin{subfigure}[b]{0.4\textwidth}
        \includegraphics[width=\textwidth]{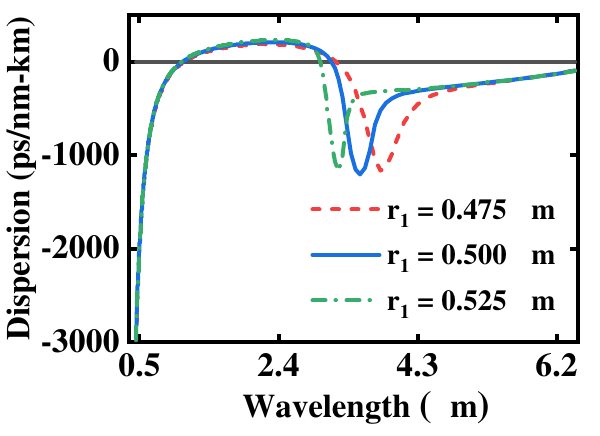}
        \caption{}
        \label{fig:tolerance-r1-v1}
        \end{subfigure}
        \begin{subfigure}[b]{0.4\textwidth}
        \includegraphics[width=\textwidth]{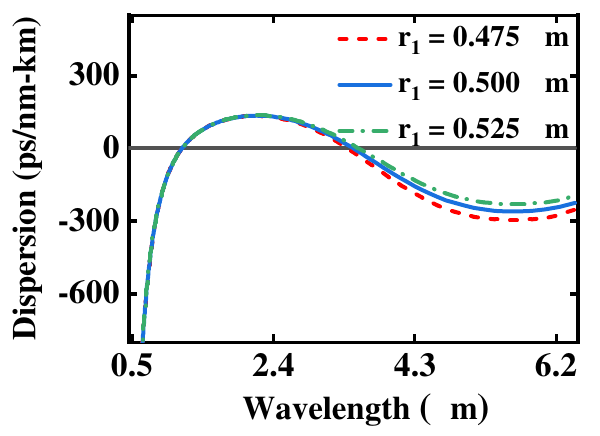}
        \caption{}
        \label{fig:tolerance-r1-v2}
        \end{subfigure}
        
        \begin{subfigure}[b]{0.4\textwidth}
        \includegraphics[width=\textwidth]{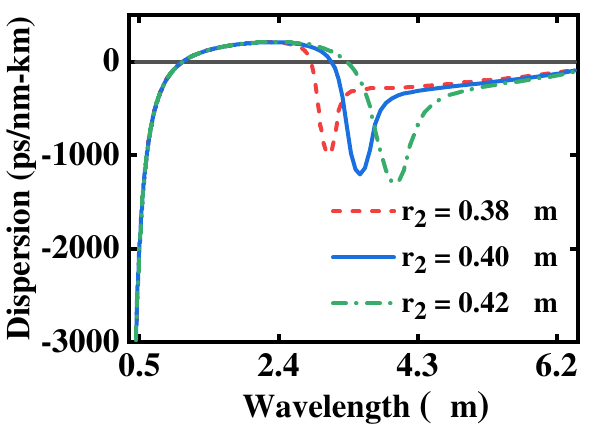}
        \caption{}
        \end{subfigure}
        \begin{subfigure}[b]{0.4\textwidth}
        \includegraphics[width=\textwidth]{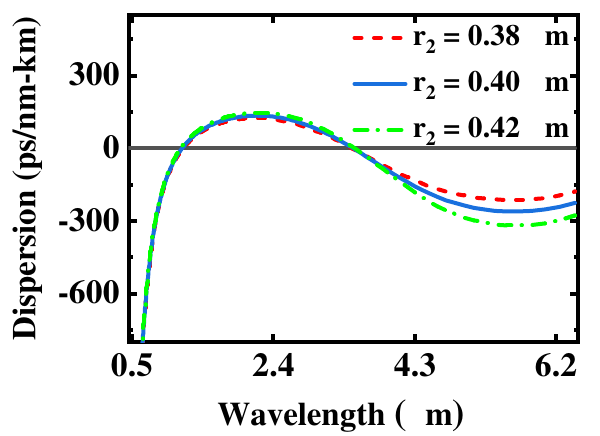}
        \caption{}
        
        \end{subfigure}
    \caption{Dispersion as function of wavelength, (a) for structure-1 when $r_1$ changes, (b) for structure-2 when $r_1$ changes, (c) for structure-1 when $r_2$ changes, and (d) for structure-2 when $r_2$ changes.}
    \label{fig:tolerance-r1-r2}
\end{figure*}
The wavelength-dependent dispersion is calculated using ~\cite{medjouri2020theoretical,karim2014dispersion},

\begin{align} 
    D(\lambda) = - \frac{\lambda}{c}\dv[2]{n_{eff}}{\lambda}.
    \label{eq:dispersion}
\end{align}

\noindent The variation in dispersion properties for both structures due to bigger air hole radius ($r_1$) and smaller air hole radius ($r_2$) is displayed in figure \ref{fig:tolerance-r1-r2}, where part (a) and part (b) show the influence of $r_1$ on the dispersion of structure-1 and structure-2, respectively.  For structure-1, air holes of radius $r_1$ are adjacent to the fiber core. This reduces the effective mode area when the value of $r_1$ increases. Also, ZDW moves from 1.1 $\mu$m to 1.07 $\mu$m  when $r_1$ varied from 0.475 $\mu$m to 0.525 $\mu$m . For structure-2, air holes of radius $r_1$ are far from the core of the fiber. As a result, it can be seen from figure \ref{fig:tolerance-r1-r2}(b) that $ r_1$ has no significant influence on the dispersion characteristics of structure-2. In table \ref{tab:my_label}, we have shown the structural dependence of ZDWs of the PCF structures. The numerical values show that ZDWs almost remained constant for 0.475 $ \mu$m, 0.50 $\mu$m, and 0.525 $\mu$m radii.
Figure \ref{fig:tolerance-r1-r2}(c) and \ref{fig:tolerance-r1-r2}(d) show the variation of dispersion due to changes of $r_2$. For structure-1, the change of $r_2$ does not have any influence on the first ZDW, but the second ZDW shifts towards the longer wavelength when the value of $r_2$ is increased. For structure-2, values of $r_2$ have a remarkable impact on the dispersion as these holes are closer to the core. The ZDW shifts towards a shorter wavelength with the increment in $r_2$.

\begin{figure*}[ht!]
    \centering
        \begin{subfigure}[b]{0.4\textwidth}
        \includegraphics[width=\textwidth]{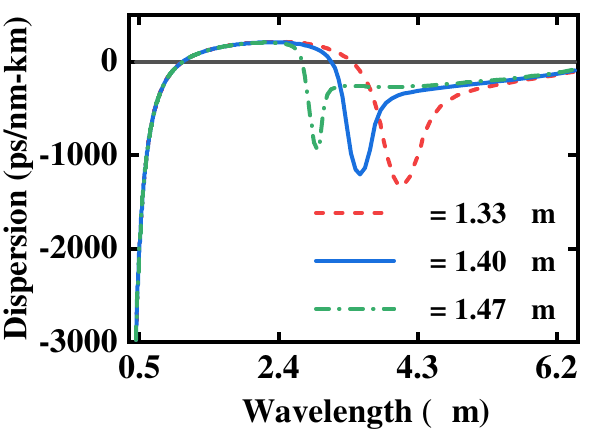}
        \caption{}
        \end{subfigure}
        \begin{subfigure}[b]{0.4\textwidth}
        \includegraphics[width=\textwidth]{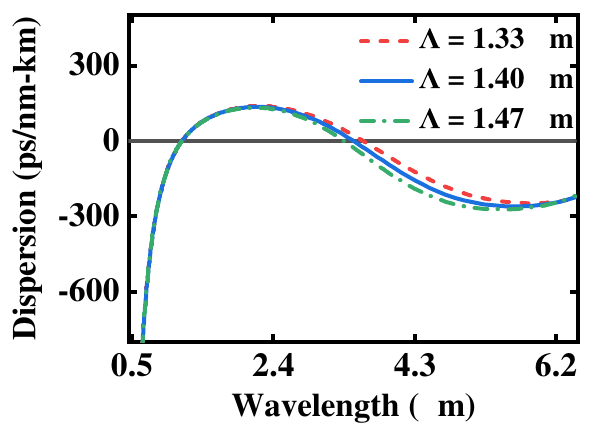}
        \caption{}
        \end{subfigure}
        
            \begin{subfigure}[b]{0.4\textwidth}
        \includegraphics[width=\textwidth]{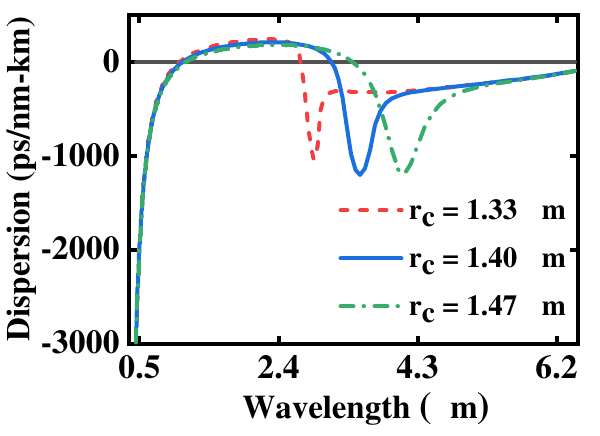}
        \caption{}
        \end{subfigure}
        \begin{subfigure}[b]{0.4\textwidth}
        \includegraphics[width=\textwidth]{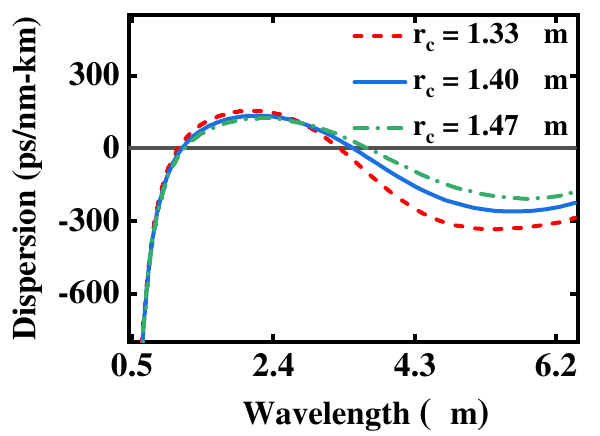}
        \caption{}
        \end{subfigure}
    
    \caption{Dispersion Characteristics due to pitch and core radius change. (a) Structure-1 due to change in $\Lambda$ (pitch). (b) Structure-2 due to change in $\Lambda$  (pitch). (c) Structure-1 due to change in core radius. (d) Structure-2 due to change in core radius.}
    \label{fig:tolerance-a-core}
\end{figure*}

\noindent We have also varied the pitch and core radius for the two structures. In figure \ref{fig:tolerance-a-core}(a) and \ref{fig:tolerance-a-core}(b), we can observe that the pitch variation changes the ZDW beyond 2.5 $\mu$m. But the first ZDWs near 1.09 $\mu$m and 1.14 $\mu$m for the two structures respectively, are very stable.
Varying the core radius also affects the dispersion characteristics. For a lower core radius of 1.33 $\mu$m, the ZDW shifts toward a shorter wavelength. For a higher core radius of 1.47 $\mu$m, the ZDW shifts toward a longer wavelength. Due to core radius change, the dispersion curve shift is high. For structure-1, as evident in figure \ref{fig:tolerance-a-core}(c), the ZDWs are 1.04 $\mu$m, 1.08 $\mu$m, 1.13 $\mu$m for core radii of 1.33 $\mu$m, 1.40 $\mu$m, 1.47 $\mu$m respectively. Again, for structure-2, as evident in figure \ref{fig:tolerance-a-core}(d) the ZDWs are 1.12 $\mu$m, 1.17 $\mu$m, 1.2 $\mu$m for core radii of 1.33 $\mu$m, 1.40 $\mu$m, and 1.47 $\mu$m, respectively.

\renewcommand{\arraystretch}{1.3}
\begin{table}
    \centering
    \caption{Study of structural Tolerance.}
    \label{tab:my_label}
    \begin{tabular}{|c|c|c |m{1.5cm}|m{1.5cm}| m{1.5cm}|m{1.5cm}|}
    \hline
    Design & Value & \% Change & \multicolumn{2}{c|}{Structure 1} & \multicolumn{2}{c|}{Structure 2}\\
    \cline{4-7}
    Parameters & ($\mu$m) & & ZDW1 & ZDW2 & ZDW1 & ZDW2 \\ \hline
       & 0.475 & $-5$ & 1.10 & 3.22 & 1.17 & 3.41 \\ \cline{2-7}
 $r_1$ & 0.500 & 0  & 1.08 & 3.15 & 1.17 & 3.50 \\
 \cline{2-7}
       & 0.525 & +5 & 1.07 & 2.98 & 1.17 & 3.59 \\ 
 \hline
       & 0.38 & $-5$ & 1.08 & 2.86 & 1.18 & 3.46 \\ \cline{2-7}
 $r_2$ & 0.40 & 0  & 1.08 & 3.15 & 1.17 & 3.50 \\ \cline{2-7}
       & 0.42 & +5 & 1.08 & 3.37 & 1.15 & 3.50 \\   
 \hline
    & 1.33 & $-5$ & 1.08 & 3.41 & 1.16 & 3.64 \\ \cline{2-7}
 $\Lambda$ & 1.40 & 0  & 1.08 & 3.15 & 1.17 & 3.50 \\ \cline{2-7}
           & 1.47 & +5 & 1.08 & 2.75 & 1.17 & 3.37 \\ 
 \hline
    & 1.33 & $-5$ & 1.04 & 2.72 & 1.12 & 3.29 \\ \cline{2-7}
$r_c$   & 1.40 & 0  & 1.08 & 3.15 & 1.17 & 3.50 \\ \cline{2-7}
        & 1.47 & +5 & 1.13 & 3.46 & 1.20 & 3.68 \\ 
 \hline
    \end{tabular}
\end{table}

\subsection{Nonlinearity Parameter}
The electrical and nuclear contributions are taken into account by the nonlinear response $R(t')$  as \cite{dudley2006supercontinuum, ALAM2021127322}

\begin{align}
    R(t') = (1-f_R)\delta(t'-t_e) + f_R\;h_R(t'),
    \label{eq:r(t)}
\end{align}

\noindent where $f_R$ is the fractional contribution of the total Raman nonlinear response. The Raman response function $h_R(t')$ is given by ~\cite{lin2006raman},

\begin{align}
        h_R(t') = \frac{\tau_1^2+\tau_2^2}{\tau_1\tau_2^2}\mathrm{exp}\left[\frac{-t'}{\tau_2}\right] \mathrm{sin}\left(\frac{t'}{\tau_1}\right).
    \label{eq:hR}
\end{align}

\noindent Here, $h_R(t')$ contains information about the vibration of the molecules as light passes through the material, for which the nonlinear parameters are $\tau_1$ = 23.1 fs, $\tau_2$ = 195 fs, and $f_R$ = 0.1 \cite{li2014low,rehan2020highly}. 
The nonlinear coefficient $\gamma$ in $\mathrm{m^{-1} W^{-1}}$ is calculated by using ~\cite{mortensen2002effective},

\begin{align}
    \gamma = \frac{2\pi n_2}{\lambda A_{eff}(\lambda)},
    \label{eq:gamma}
\end{align}

\noindent where, $n_2 = 6.8 \times 10^{-18}$ $\mathrm{m^2 W^{-1}}$ is the nonlinear refractive index of the material. The effective mode area can be computed by using mode field distribution over the cross-section as \cite{rehan2020highly},

\begin{align}
    A_{eff}(\lambda) = \frac{(\int\int_{-\infty}^{\infty}|E|^2\,dx\,dy)^2}{(\int\int_{-\infty}^{\infty}|E|^4\,dx\,dy)}.
    \label{eq:Aeff}
\end{align}

\begin{figure}[!ht]
    \centering
    \includegraphics[width= 12 cm]{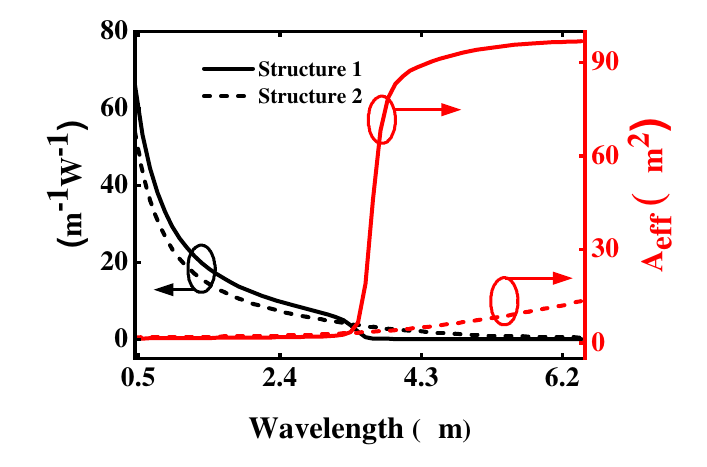}

    \hfill
    \vspace{-2.5em}
    \caption{$A_{eff}$ and $\gamma$ for both the structures.}
    \label{fig:Aeff_gamma}
\end{figure}

\noindent The Variation of effective mode area $(A_{eff})$ and nonlinear coefficient $(\gamma)$ with respect to wavelength for both the structures are illustrated in figure \ref{fig:Aeff_gamma}. The  $A_eff$ for structure-1 and structure-2 at 1.2 $\mu$m wavelength are 1.5939 $\mathrm{\mu m^2}$  and 20.0313 $\mathrm{\mu m^2}$ respectively. Also, we have found the nonlinear coefficient for  structure-1 and structure-2 at 1.2 $\mu$m wavelength as 22.01 $\mathrm{W^{-1}m^{-1}}$ and 17.99202 $\mathrm{W^{-1}m^{-1}}$ respectively. The higher values of $\gamma$ are clearly a sign of better design and a good choice of material.  Structure-1 has the lower effective mode area, hence nonlinear coefficient for this structure will be higher as shown in the figure.

\section{Supercontinuum Generation (SCG)}
In this section, the investigations of SCG in the highly nonlinear PCFs are presented. The generation of SC is governed by nonlinear Schrödinger equation (NLSE) ~\cite{dudley2002supercontinuum},

\begin{align}
    \frac{\partial A(z,t)}{\partial z}+\frac{\alpha}{2}A(z,t) - \sum_{k \geq 2} i^{k+1} \frac{\beta_k}{k!} \frac{\partial^{k} A(z,t)}{\partial t^{k}} = i\gamma \left (1 + \frac{i}{\omega_0} \frac{\partial}{\partial t}\right) \left [A(z,t) \int_{-\infty}^{t} R(t^{'}) |A(z,t-t^{'})|^2 dt^{'} \right]
    \label{eq:scg}
\end{align}

\noindent The left side of  (\ref{eq:scg}) describes the linear propagation effects with $\alpha$ and $\beta_k$ denoting the linear loss and dispersion coefficients, respectively. As both structures have a length of 25 mm only, the loss is not expected to impact  SC significantly. For this reason, we have considered loss as zero during the analysis. 
Here, $A = A(z, t)$ is the envelope of electric ﬁeld. Again, the right side of  (\ref{eq:scg}) signifies the nonlinear effects with nonlinear coefficient $ \gamma $ which is associated with nonlinear refractive index and effective modal area.
The NLSE is solved numerically using the split-step Fourier method which is popular for its simplicity and accuracy. $\beta$ can be  expanded by  using the Taylor Series ~\cite{rehan2020highly},

\begin{align}
    \beta(\omega) = \beta(\omega_0) + \beta_1(\omega_0)(\omega-\omega_0) + \frac{1}{2!} \beta_2(\omega_0)(\omega-\omega_0)^2+...
    \label{eq:taylor}
\end{align}

\noindent In (\ref{eq:taylor}) $\beta_n$ is the nth dispersion coefficient which is calculated by $\beta_n = \Big ( \frac{d^n\beta (\omega)}{d\omega^n} \Big )_{\omega = \omega_0} $. In this work, we have considered dispersion parameters ($\beta$) values up to order 10 to ensure accuracy, and the values are given in table \ref{table:beta_s1}.

\begin{table}[!ht]
\renewcommand{\arraystretch}{1.2}
\centering
\caption{Dispersion parameters for both the structures.}
\label{table:beta_s1}
\begin{minipage}[t]{1\linewidth}
    \centering
    \begin{tabular}{ c c c }
    \hline
     & Structure 1 & Structure 2 \\
     \hline
     $\beta_2(\,\mathrm{ps^2/km})$ & $-4.3372 \times 10^1$ & $ -1.0307 \times 10^1$ \\
     
     $\beta_3(\,\mathrm{ps^3/km})$ &  $2.9063 \times 10^{-1}$ &  $2.3046 \times 10^{-1}$\\
     
     $\beta_4(\,\mathrm{ps^4/km})$ & $-5.0112 \times 10^{-4}$ & $-3.5315\times 10^{-4}$ \\ 
     
     $\beta_5(\,\mathrm{ps^5/km})$ & $ 1.4268 \times 10^{-6}$ & $9.5215 \times 10^{-7}$\\ 
     
     $\beta_6(\,\mathrm{ps^6/km})$ & $-4.5986 \times 10^{-9}$ & $-2.7127\times 10^{-9}$ \\
     
     $\beta_7(\,\mathrm{ps^7/km})$ & $1.6898 \times 10^{-11}$& $8.7043 \times 10^{-12}$ \\ 
     
     $\beta_8(\,\mathrm{ps^8/km})$ & $1.9995 \times 10^{-14}$ & $1.9631 \times 10^{-13}$ \\
     
     $\beta_9(\,\mathrm{ps^9/km})$ & $-6.3870 \times 10^{-15}$ & $-2.2423 \times 10^{-14}$ \\ 
     
     $\beta_{10}(\,\mathrm{ps^{10}/km})$ & $-2.0522 \times 10^{-15}$ & $-4.3991 \times 10^{-15}$\\
     \hline
    \end{tabular}    
\end{minipage}

\end{table}

\noindent We have used a hyperbolic secant pulse of 50 fs full width at half maximum (FWHM) and 0.5 kW peak power which is expressed by $A(z=0,t) = \sqrt{P_0}\mathrm{sech}\left(\frac{t}{t_0}\right)$. Here, $t_0 = T_{FWHM}/1.7627$ and $P_0$ is the peak power of the pulse. Reports of this kind of InGaAs laser pulse are available in literature~\cite{tansu2003high}. We set the pump wavelength at 1200 nm (250 THz) in the anomalous dispersion regime which is very close to the ZDW. In an anomalous regime, spectral broadening occurs due to an initial small amount of self-phase modulation (SPM) and soliton fissions which causes the explosive broadening of spectra. 

\begin{figure*}[ht!]
    \centering
        \begin{subfigure}[b]{0.4\textwidth}
        \includegraphics[width=\textwidth]{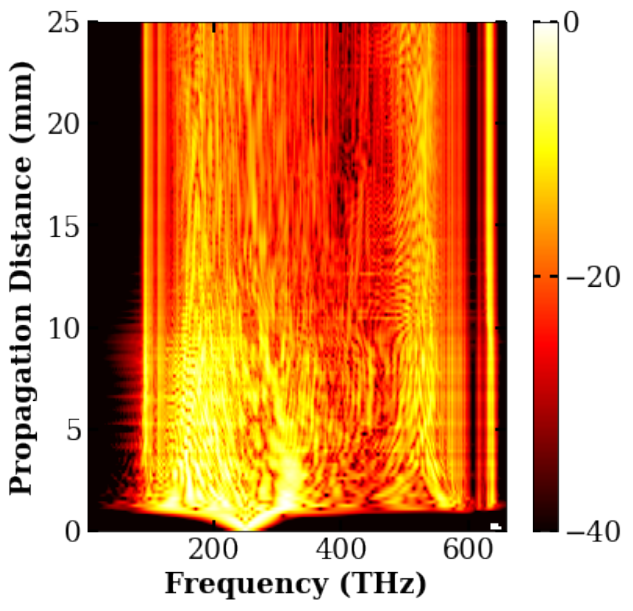}
        \caption{}
        \end{subfigure}
        \begin{subfigure}[b]{0.4\textwidth}
        \includegraphics[width=\textwidth]{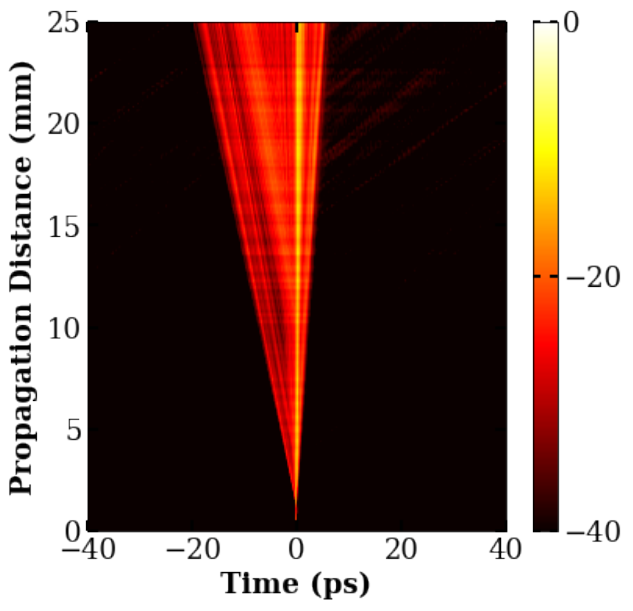}
        \caption{}
        \end{subfigure}
        
        \begin{subfigure}[b]{0.4\textwidth}
        \includegraphics[width=\textwidth]{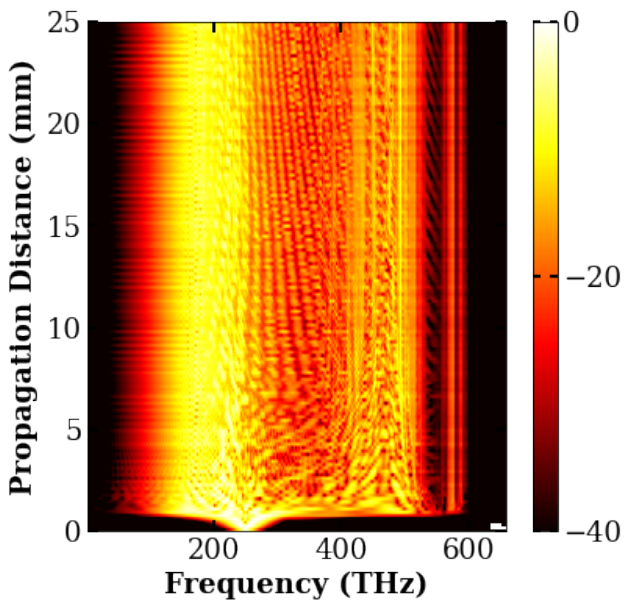}
        \caption{}
        \end{subfigure}
        \begin{subfigure}[b]{0.4\textwidth}
        \includegraphics[width=\textwidth]{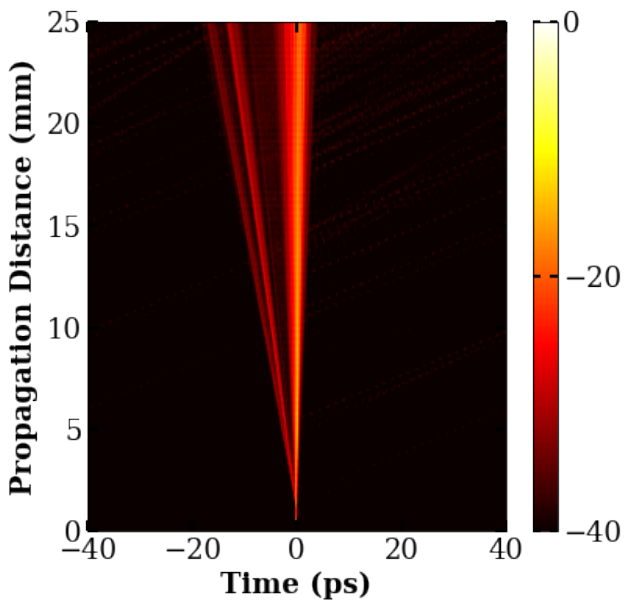}
        \caption{}
        \end{subfigure}
    
    \caption{Spectral and temporal evolutions at various propagation distances for pulse power of 0.5 kW and FWHM of 50 fs, (a)-(b) structure-1, (c)-(d) structure-2.}
    \label{fig:evolution}
\end{figure*}

\noindent Figure \ref{fig:evolution} depicts the spectral and temporal evolutions of input pulse with propagation distance for both structures. We can see that most of the broadening happens within a few millimeters. The characteristic lengths of both fibers are calculated using ~\cite{dudley2006supercontinuum},

\begin{align}
    N^2 = \frac{L_D}{L_{NL}}, \hspace{3mm} 
    L_D = \frac{t_0^2}{|\beta_2|},\hspace{3mm} 
    L_{NL} = \frac{1}{\gamma P_0}, \hspace{3mm}
    L_{fiss} = \frac{L_D}{N}.
    \label{eq:characteristics-length}
\end{align}

\begin{figure*}[ht!]
    \centering
        \begin{subfigure}[b]{0.42\textwidth}
        \includegraphics[width=\textwidth]{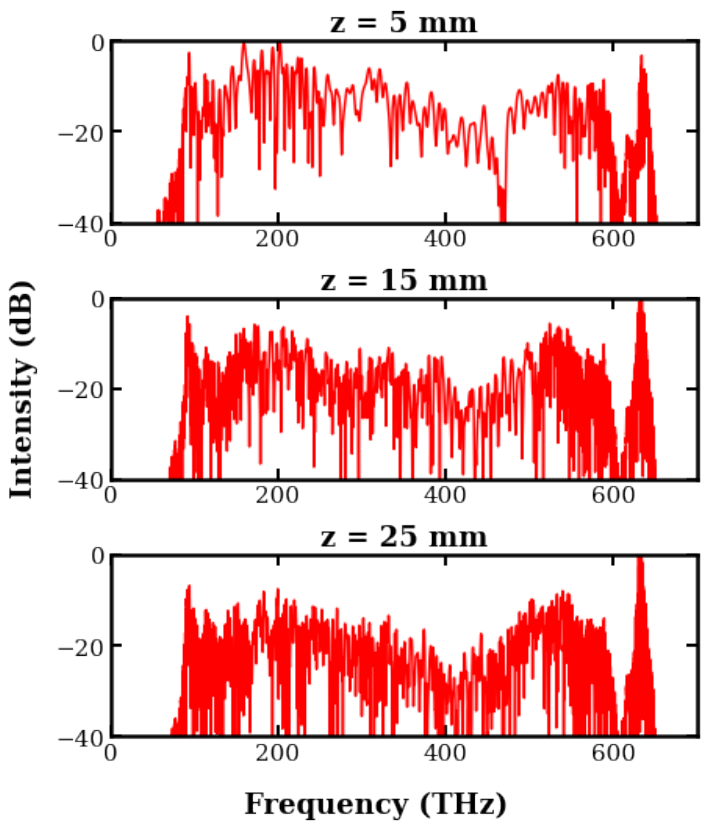}
        \caption{}
        \end{subfigure}
        \begin{subfigure}[b]{0.42\textwidth}
        \includegraphics[width=\textwidth]{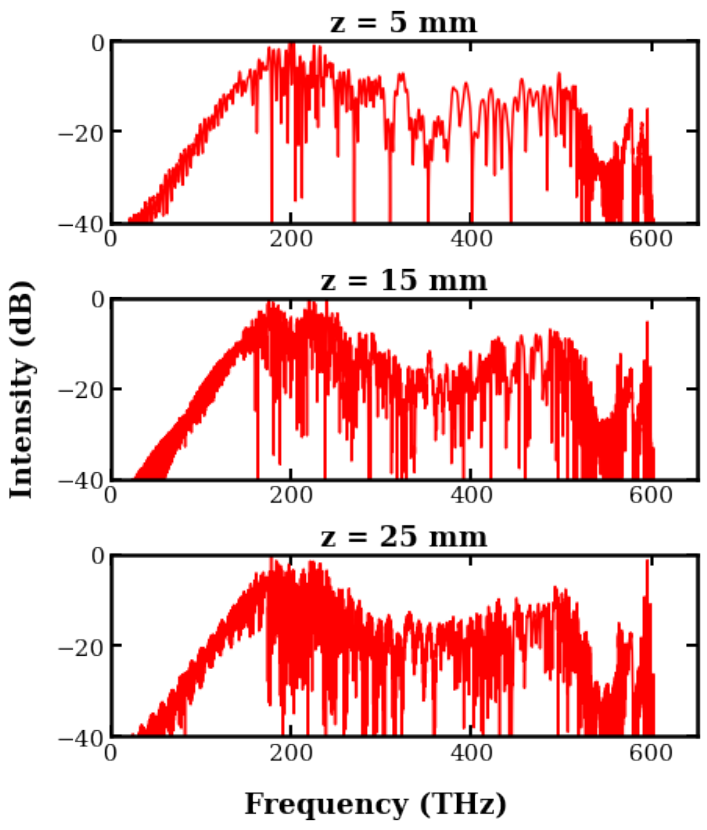}
        \caption{}
        \end{subfigure}
    
    \caption{Power spectral density (PSD) at different distances, (a) Structure-1. (b) Structure-2.}
    \label{fig:psd}
\end{figure*}

\noindent Here, $L_D$, $L_{NL}$, $L_{fiss}$ denote the dispersive, nonlinear, and soliton fission lengths, respectively. 
In this study, we have obtained: $L_D$ = 18.55 mm, $L_{NL}$ = 0.09 mm, $L_{fiss}$ = 1.29 mm for structure-1 and $L_D$ = 61.56 mm, $L_{NL}$ = 0.11 mm, $L_{fiss}$ = 2.616 mm for structure-2.
So, SPM initiates the spectral broadening and then soliton fission follows as the input pulse transforms into a series of fundamental solitons due to a small perturbation, such as high-order dispersion, or SRS. The solitons shift toward the lower frequency due to self-frequency shifting with further propagation along the fiber. This leads to an expansion in the left side of the spectrum. This effect can be visualized as the right side of the spectrum remains almost the same but the left side changes due to soliton dynamics. Figure \ref{fig:psd}(a) and \ref{fig:psd}(b) show the normalized spectral intensity versus frequency at different fiber lengths which verify that both structures are capable of generating flat ultra-broadband SCs. Also, we can see that the spectra of SC are extended from 56.06 THz (5.3  $\mu$m) to 666.66 THz (0.45 $\mu$m) for structure-1 and from 46.25 THz (6.5 $\mu$m) to 625.0 THz (0.48 $\mu$m) for structure-2.

\begin{figure}[!ht]
    \centering
    \includegraphics[width=0.9\textwidth]{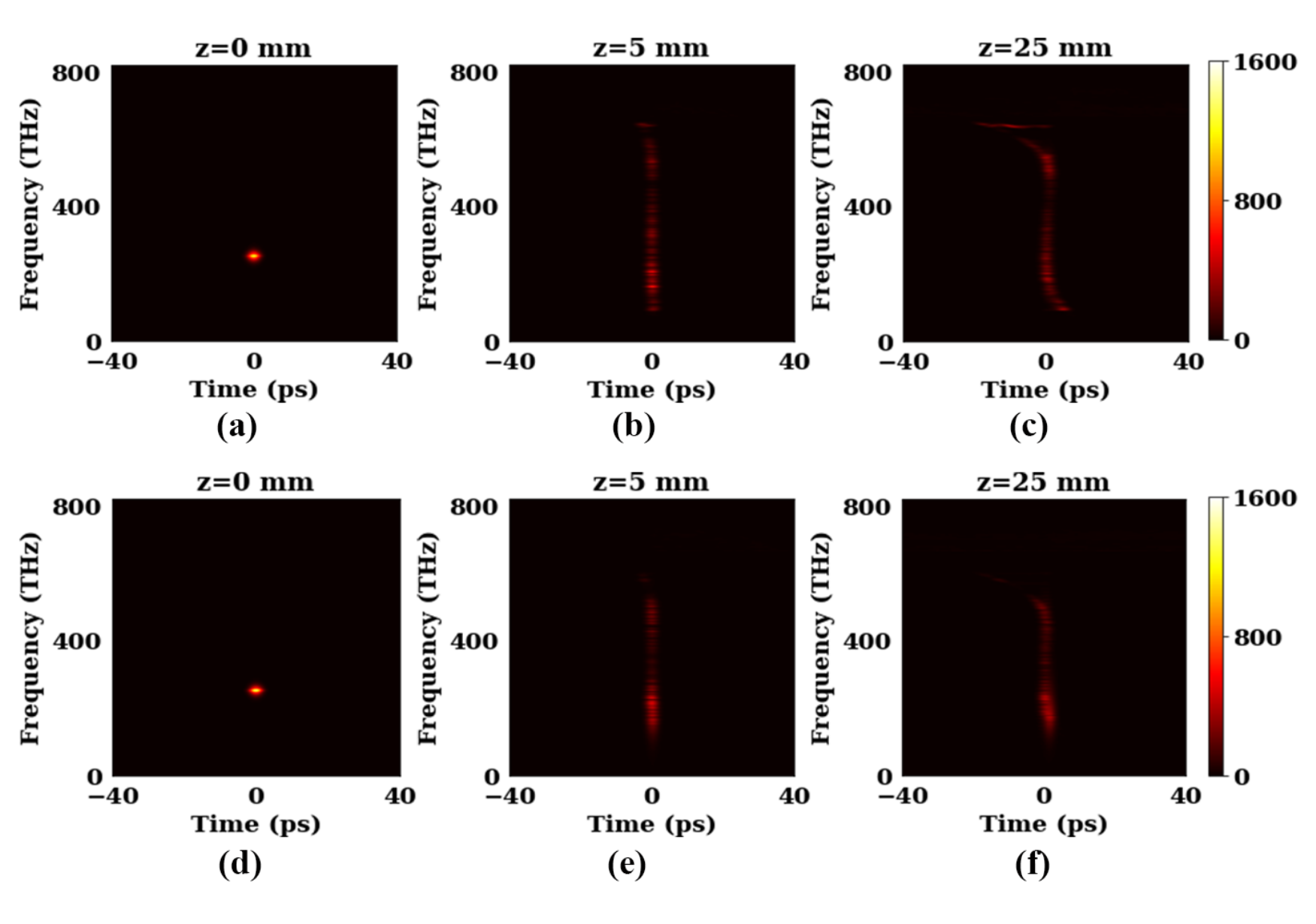}
    \hfill
    \caption{Spectrogram at various propagation distances, (a)-(b)-(c) Structure-1, (d)-(e)-(f) Structure-2.}
    \label{fig:spectrogram1}
\end{figure}

\noindent When the pulse propagates, it experiences more nonlinear activity, which makes the spectra more wavelength sensitive and oscillation grows as the propagation distance increases \cite{kalantari2018ultra,li2014low}. For a more detailed analysis, we have presented spectrograms at three different fiber lengths in figure \ref{fig:spectrogram1}. This figure shows that initially at 0 mm fiber length, only a single frequency component exists at the pumping wavelength. Then, in a very small fraction of time and at only 5 mm fiber length, a large number of frequency components are generated as a result of various nonlinear processes. With further propagation, more frequency components are generated and extended supercontinuums are produced.

\begin{figure}[!ht]
    \centering
    \includegraphics[width= 9 cm]{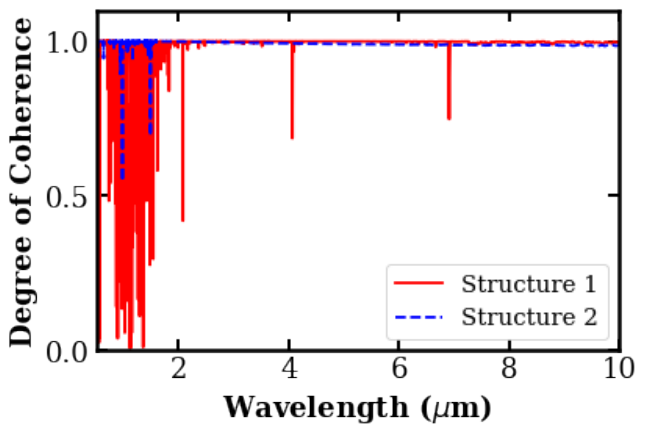}
    \hfill
    \caption{Degree of Coherence for over 25 mm long PCFs.}
    \label{fig:degree_of_coherence}
\end{figure}

\noindent We have also calculated the degree of coherence numerically by using \cite{dudley2006supercontinuum},

\begin{align}
   \big |g_{12}^{(1)}(\lambda,t_{1} - t_{2}) \big | = \Bigg| \frac{ \langle E_{1}^{*}(\lambda,t_{1})E_{2}(\lambda,t_{2}) \rangle }{\sqrt{\langle |E_{1}(\lambda,t_{1})|^{2}\rangle \langle|E_{2}(\lambda,t_{2})|^{2}\rangle}}  \Bigg |.
   \label{eq:coherence}
\end{align}
Here, $E_1(\lambda,t)$ and $E_2(\lambda,t)$ are independently generated pairs of SC which are emitted at different times $t_1$ and $t_2$. The angular brackets denote an ensemble average.  Here, $\big |g_{12}^{(1)} \big |$ is a measure of amplitude which is considered as the first-order degree of coherence as a function of wavelength. The values of $\big |g_{12}^{(1)} \big |$ are positive  from 0 to 1. Value 1 indicates the perfect spectral coherence of the source and value 0 corresponds to complete decoherence. In an anomalous dispersion regime, the modulation instability (MI) that arises from the high peak power input noise can make the SC sources incoherent. The MI is an amplification process that creates sidebands in the SC spectra and rapidly amplifies the sidebands. As we have used a low peak power pulse as input, the MI  does not affect the SC spectra which can lead to a coherent source. Figure \ref{fig:degree_of_coherence} shows the analysis of coherence for 25 mm long fibers with the designs of structure-1 and structure-2. The values of coherence for both structures are very close to 1. So, it can be noted that both structures are capable of generating coherent SCs. 

Finally, table \ref{table:SC_compare} shows a comparison between SC generated in our designs and other recent PCF designs where various nonlinear materials are used. H. Saghaei \textit{et. al.} \cite{saghaei2015midinfrared} and M. R. Karim \textit{et. al.} \cite{karim2017all} used $As_2Se_3$ and $Ge_{11.5}As_{24}Se_{64.5}$ chalcogenide materials to achieve high bandwidth of 3.86 $\mathrm{\mu}$m and 3.5 $\mathrm{\mu}$m, respectively, however they employed high peak power pulses of 10 kW and 5 kW.

\renewcommand{\arraystretch}{1.2}
\begin{table}[ht!]
\centering
\caption{Comparison of generated SCs in various PCF designs.}
\label{table:SC_compare}
\begin{minipage}[t]{1\linewidth}
    \centering
    \begin{tabular}{>{\centering\arraybackslash}p{2.5cm}  >{\centering\arraybackslash}p{3cm} >{\centering\arraybackslash}p{1.5cm} >{\centering\arraybackslash}p{2.5cm} >{\centering\arraybackslash}p{2cm}}
    \hline
      Material & Pump Wavelength ($\mu$m)  & Peak Power (kW) & SC bandwidth ($\mu$m)& Reference\\
     \hline
     $SiO_2$  & 1.3 & 1.38 & 1.1 - 1.7& \cite{FERHAT2018106}\\
     $As_2Se_3$ & 4.6 & 10 & 3 - 6.86 & \cite{saghaei2015midinfrared} \\
     $AsSe_2/As_2S_5$ & 2.7 & 0.095 & 2.2 - 3.3 &\cite{liu2016coherent}\\
     $Ge_{11.5}As_{24}Se_{64.5}$ &3.1 &5 & 2 - 5.5 & \cite{karim2017all}\\
     $As_2Se_3$ & 4.3945 & 4.25 &
     3.866 - 5.958 & \cite{wang2019flattened} \\
     $Ge_{20}Sb_{15}Se_{65}$ (structure-1) & 1.2 & 0.5 & 0.45 - 5.3 & This work\\
     $Ge_{20}Sb_{15}Se_{65}$ (structure-2) & 1.2 & 0.5 & 0.48 - 6.5 & This work\\
     \hline
    \end{tabular}
    
\end{minipage}
\end{table}
M. L. Ferhat \textit{et. al.} \cite{FERHAT2018106} and L. Liu \textit{et. al.} \cite{liu2016coherent} employed relatively low peak power, though SC bandwidth was limited to 0.6 $\mathrm{\mu}$m and 1.1 $\mathrm{\mu}$m respectively. 
However, SC with a comparatively higher bandwidth of 2.092 $\mathrm{\mu}$m was reported by E. Wang \textit{et. al.} \cite{wang2019flattened}, but the peak power of the pulse was comparatively high (4.25 kW) and the pumping wavelength was longer (4.3945 $\mathrm{\mu}$m). On the contrary, the designed structures in this work produce broadband (from visible to mid-IR) SCs with just 0.5 kW peak power and 1.2 $\mathrm{\mu}$m pumping wavelength, which is a major improvement over previous designs. Both the designs allow to pump at 1.2 $\mu$m as they provide zero dispersion wavelengths (ZDW) at shorter wavelengths. Besides these, the designs have higher nonlinear coefficients which indicate a higher nonlinear action to broaden the spectrum.  A comparison between these two structures shows that Structure 2 has a wider SC spectrum and less confinement loss than Structure 1, and hence may be superior to Structure 1. Additionally, structure 2 shows a more uniform and stable degree of coherence.

\section{Conclusion}
We have developed two SC sources in this work by using two new PCF designs with chalcogenide material. The structures were capable of confining fundamental modes in the core region, resulting in minimal confinement losses. Again, higher values of nonlinear coefficients (22.01 $\mathrm{W^{-1}m^{-1}}$ for structure-1 and 17.99202 $\mathrm{W^{-1}m^{-1}}$ for structure-2) and almost flat dispersion characteristics were found for both fibers. We have also conducted a tolerance study to account for any possible variation in structural parameters resulting from fabrication constraints. For solving NLSE, we have used the split-step Fourier method (SSFM) and by launching a 50 fs hyperbolic secant pulse of only 0.5 kW into the fibers, we have generated supercontinuums ranging from 0.45 $\mu$m to 5.3 $\mu$m (structure-1) and 0.48 $\mu$m to 6.5 $\mu$m (structure-2). Thus, the proposed designs hold promise for applications in the fields of telecommunication, bio-sensing, medical imaging, and spectroscopy.

\section*{References}
\bibliographystyle{unsrt}
\bibliography{bib.bib}

\begin{thebibliography}{10}

\bibitem{dudley2006supercontinuum}
J.~M. Dudley, G.~Genty, and S.~Coen.
\newblock Supercontinuum generation in photonic crystal fiber.
\newblock {\em Rev. Mod. Phys.}, 78:1135--1184, Oct 2006.

\bibitem{ke2009mid}
K.~Ke, C.~Xia, M.~N. Islam, M.~J. Welsh, and M.~J. Freeman.
\newblock Mid-infrared absorption spectroscopy and differential damage in vitro between lipids and proteins by an all-fiber-integrated supercontinuum laser.
\newblock {\em Opt. Express}, 17(15):12627--12640, Jul 2009.

\bibitem{marks2002study}
D.~L. Marks, A.~L. Oldenburg, J.~J. Reynolds, and S.~A. Boppart.
\newblock Study of an ultrahigh-numerical-aperture fiber continuum generation source for optical coherence tomography.
\newblock {\em Opt. Lett.}, 27(22):2010--2012, Nov 2002.

\bibitem{dupont2012ir}
S.~Dupont, C.~Petersen, J.~Th{\o}gersen, C.~Agger, O.~Bang, and S.~R. Keiding.
\newblock Ir microscopy utilizing intense supercontinuum light source.
\newblock {\em Opt. Express}, 20(5):4887--4892, Feb 2012.

\bibitem{petersen2018mid}
C.~R. Petersen, N.~Prtljaga, M.~Farries, J.~Ward, B.~Napier, G.~R. Lloyd, J.~Nallala, N.~Stone, and O.~Bang.
\newblock Mid-infrared multispectral tissue imaging using a chalcogenide fiber supercontinuum source.
\newblock {\em Opt. Lett.}, 43(5):999--1002, Mar 2018.

\bibitem{boyraz200010}
O.~Boyraz, J.~Kim, M.~N. Islam, E.~Coppinger, and B.~Jalali.
\newblock 10 \uppercase{g}b/s multiple wavelength,coherent short pulse source based on spectral carving of supercontinuum generated in fibers.
\newblock {\em J. Lightwave Technol.}, 18(12):2167, Dec 2000.

\bibitem{Ohara06}
T.~Ohara, H.~Takara, T.~Yamamoto, H.~Masuda, T.~Morioka, M.~Abe, and H.~Takahashi.
\newblock Over-1000-channel ultradense \uppercase{WDM} transmission with supercontinuum multicarrier source.
\newblock {\em J. Lightwave Technol.}, 24(6):2311, Jun 2006.

\bibitem{PhysRevLett.24.592}
R.~R. Alfano and S.~L. Shapiro.
\newblock Observation of self-phase modulation and small-scale filaments in crystals and glasses.
\newblock {\em Phys. Rev. Lett.}, 24:592--594, Mar 1970.

\bibitem{PhysRevLett.24.584}
R.~R. Alfano and S.~L. Shapiro.
\newblock Emission in the region 4000 to 7000 \uppercase{\aa} via four-photon coupling in glass.
\newblock {\em Phys. Rev. Lett.}, 24:584--587, Mar 1970.

\bibitem{begum2019near}
F.~Begum and P.~E. Abas.
\newblock Near infrared supercontinuum generation in silica based photonic crystal fiber.
\newblock {\em Progress in Electromagnetics Research}, 89:149--159, 2019.

\bibitem{FERHAT2018106}
M.~L. Ferhat, L.~Cherbi, and I.~Haddouche.
\newblock Supercontinuum generation in silica photonic crystal fiber at 1.3 $\mathrm{\mu}$m and 1.65 $\mathrm{\mu}$m wavelengths for optical coherence tomography.
\newblock {\em Optik}, 152:106--115, 2018.

\bibitem{Jiang2015}
X.~Jiang, N.~Y. Joly, M.~A. Finger, F.~Babic, G.~K.~L. Wong, J.~C. Travers, and P.~S.~J. Russell.
\newblock Deep-ultraviolet to mid-infrared supercontinuum generated in solid-core \uppercase{ZBLAN} photonic crystal fibre.
\newblock {\em Nature Photonics}, 9(2):133--139, Feb 2015.

\bibitem{chauhan2019dispersion}
P.~Chauhan, A.~Kumar, and Y.~Kalra.
\newblock A dispersion engineered silica-based photonic crystal fiber for supercontinuum generation in near-infrared wavelength region.
\newblock {\em Optik}, 187:230--237, 2019.

\bibitem{tarnowski2016coherent}
K.~Tarnowski, T.~Martynkien, P.~Mergo, K.~Poturaj, G.~Sobo\'{n}, and W.~Urba\'{n}czyk.
\newblock Coherent supercontinuum generation up to 2.2 $\mathbf{\mu}$m in an all-normal dispersion microstructured silica fiber.
\newblock {\em Opt. Express}, 24(26):30523--30536, Dec 2016.

\bibitem{zakery2007optical}
A.~Zakery and S.~R. Elliott.
\newblock {\em {Optical nonlinearities in chalcogenide glasses and their applications}}.
\newblock Springer, 2007.

\bibitem{jayasuriya2019mid}
D.~Jayasuriya, C.~R. Petersen, D.~Furniss, C.~Markos, Z.~Tang, M.~S. Habib, O.~Bang, T.~M. Benson, and A.~B. Seddon.
\newblock Mid-ir supercontinuum generation in birefringent, low loss, ultra-high numerical aperture \uppercase{g}e-\uppercase{A}s-\uppercase{s}e-\uppercase{t}e chalcogenide step-index fiber.
\newblock {\em Opt. Mater. Express}, 9(6):2617--2629, Jun 2019.

\bibitem{wang2019flattened}
E.~Wang, J.~Li, J.~Li, Q.~Cheng, X.~Zhou, and H.~Jiang.
\newblock Flattened and broadband mid-infrared super-continuum generation in \uppercase{a}$\mathrm{s_2}$\uppercase{s}$\mathrm{e_3}$ based holey fiber.
\newblock {\em Optical and Quantum Electronics}, 51(1):10, Dec 2018.

\bibitem{saghaei2015ultra}
H.~Saghaei, M.~K. Moravvej-Farshi, M.~Ebnali-Heidari, and M.~N. Moghadasi.
\newblock Ultra-wide mid-infrared supercontinuum generation in \uppercase{A}$\mathrm{s_{40}}$ \uppercase{s}$\mathrm{e_{60}}$ chalcogenide fibers: Solid core \uppercase{PCF} versus \uppercase{SIF}.
\newblock {\em IEEE Journal of Selected Topics in Quantum Electronics}, 22(2):279--286, 2016.

\bibitem{ghosh2019chalcogenide}
A.~N. Ghosh, M.~Meneghetti, C.~R. Petersen, O.~Bang, L.~Brilland, S.~Venck, J.~Troles, J.~M. Dudley, and T.~Sylvestre.
\newblock Chalcogenide-glass polarization-maintaining photonic crystal fiber for mid-infrared supercontinuum generation.
\newblock {\em Journal of Physics: Photonics}, 1(4):044003, sep 2019.

\bibitem{chaitanya2016ultra}
A.~Chaitanya, T.~S. Saini, A.~Kumar, and R.~K. Sinha.
\newblock Ultra broadband mid-\uppercase{ir} supercontinuum generation in \uppercase{g}$\mathrm{e_{11.5}}$\uppercase{a}$\mathrm{s_{24}}$\uppercase{s}$\mathrm{e_{64.5}}$ based chalcogenide graded-index photonic crystal fiber: design and analysis.
\newblock {\em Applied Optics}, 55(36):10138--10145, 2016.

\bibitem{zhanqiang2018mid}
H.~Zhanqiang, Y.~Zhang, H.~Zhou, Z.~Wang, and X.~Zeng.
\newblock Mid-infrared high birefringence bow-tie-type \uppercase{g}$\mathrm{e_{20}}$\uppercase{s}$\mathrm{b_{15}}$\uppercase{s}$\mathrm{e_{65}}$ based \uppercase{PCF} with large nonlinearity by using hexagonal elliptical air hole.
\newblock {\em Fiber and Integrated Optics}, 37(1):21--36, 2018.

\bibitem{amiri2019design}
I.~S. Amiri, M.~A. Khalek, S.~Chakma, B.~K. Paul, K.~Ahmed, V.~Dhasarathan, and M.~{Mani Rajan}.
\newblock Design of \uppercase{g}$\mathrm{e_{20}}$\uppercase{s}$\mathrm{b_{15}}$\uppercase{s}$\mathrm{e_{65}}$ embedded rectangular slotted quasi photonic crystal fiber for higher nonlinearity applications.
\newblock {\em Optik}, 184:63--69, 2019.

\bibitem{Zhu:02}
Z.~Zhu and T.~G. Brown.
\newblock Full-vectorial finite-difference analysis of microstructured optical fibers.
\newblock {\em Opt. Express}, 10(17):853--864, Aug 2002.

\bibitem{Nielsen:03}
M.~D. Nielsen and N.~A. Mortensen.
\newblock Photonic crystal fiber design based on the v-parameter.
\newblock {\em Opt. Express}, 11(21):2762--2768, Oct 2003.

\bibitem{jurgensen1975dispersion}
K.~J\"{u}rgensen.
\newblock Dispersion-optimized optical single-mode glass fiber waveguides.
\newblock {\em Appl. Opt.}, 14(1):163--168, Jan 1975.

\bibitem{senior2009optical}
J.~M. Senior.
\newblock {\em Optical fiber communications: principles and practice}.
\newblock Pearson Education, 2009.

\bibitem{ALAM2021127322}
M.~Z. Alam, M.~I. Tahmid, S.~T. Mouna, M.~A. Islam, and M.~S. Alam.
\newblock Design of a novel star type photonic crystal fiber for mid-infrared supercontinuum generation.
\newblock {\em Optics Communications}, 500:127322, 2021.

\bibitem{Zahid_JNP}
Z.~Rahman, M.~A. Rahman, M.~A. Islam, and M.~S. Alam.
\newblock {Design of an elliptical air-hole dual-core photonic crystal fiber for over two octaves spanning supercontinuum generation}.
\newblock {\em Journal of Nanophotonics}, 13(4):1 -- 12, 2019.

\bibitem{kalantari2018ultra}
M.~Kalantari, A.~Karimkhani, and H.~Saghaei.
\newblock Ultra-wide mid-ir supercontinuum generation in \uppercase{a}$s_{2}$$\uppercase{s_3}$ photonic crystal fiber by rods filling technique.
\newblock {\em Optik}, 158:142--151, 2018.

\bibitem{medjouri2020theoretical}
A.~Medjouri and D.~Abed.
\newblock Theoretical study of coherent supercontinuum generation in chalcohalide glass photonic crystal fiber.
\newblock {\em Optik}, 219:165178, 2020.

\bibitem{karim2014dispersion}
M.~R. Karim, B.~M.~A. Rahman, and G.~P. Agrawal.
\newblock Dispersion engineered \uppercase{g}$\mathrm{e_{11.5}}$\uppercase{a}$\mathrm{s_{24}}$\uppercase{s}$\mathrm{e_{64.5}}$ nanowire for supercontinuum generation: A parametric study.
\newblock {\em Opt. Express}, 22(25):31029--31040, Dec 2014.

\bibitem{lin2006raman}
Q.~Lin and G.~P. Agrawal.
\newblock Raman response function for silica fibers.
\newblock {\em Opt. Lett.}, 31(21):3086--3088, Nov 2006.

\bibitem{li2014low}
X.~Li, W.~Chen, T.~Xue, J.~Gao, W.~Gao, L.~Hu, and M.~Liao.
\newblock Low threshold mid-infrared supercontinuum generation in short fluoride-chalcogenide multimaterial fibers.
\newblock {\em Opt. Express}, 22(20):24179--24191, Oct 2014.

\bibitem{rehan2020highly}
Anjali, M.~Rehan, G.~Kumar, and V.~Rastogi.
\newblock Highly nonlinear silica spiral photonic crystal fiber for supercontinuuam generation extending from 320 nm to 2620 nm.
\newblock {\em Optik}, 208:163897, 2020.

\bibitem{mortensen2002effective}
N.~A. Mortensen.
\newblock Effective area of photonic crystal fibers.
\newblock {\em Opt. Express}, 10(7):341--348, Apr 2002.

\bibitem{dudley2002supercontinuum}
J.~M. Dudley, L.~Provino, N.~Grossard, H.~Maillotte, R.~S. Windeler, B.~J. Eggleton, and S.~Coen.
\newblock Supercontinuum generation in air--silica microstructured fibers with nanosecond and femtosecond pulse pumping.
\newblock {\em JOSA B}, 19(4):765--771, 2002.

\bibitem{tansu2003high}
N.~Tansu, J.-Y. Yeh, and L.~J. Mawst.
\newblock High-performance 1200-nm \uppercase{i}n\uppercase{g}a\uppercase{a}s and 1300-nm \uppercase{i}n\uppercase{g}a\uppercase{a}s quantum-well lasers by metalorganic chemical vapor deposition.
\newblock {\em IEEE Journal of Selected Topics in Quantum Electronics}, 9(5):1220--1227, 2003.

\bibitem{saghaei2015midinfrared}
H.~Saghaei, M.~Ebnali-Heidari, and M.~K. Moravvej-Farshi.
\newblock Midinfrared supercontinuum generation via \uppercase{a}$\mathrm{s_2}$\uppercase{s}$\mathrm{e_3}$ chalcogenide photonic crystal fibers.
\newblock {\em Appl. Opt.}, 54(8):2072--2079, Mar 2015.

\bibitem{karim2017all}
M.~R. Karim, H.~Ahmad, and B.~M.~A. Rahman.
\newblock All-normal dispersion chalcogenide \uppercase{pcf} for ultraflat mid-infrared supercontinuum generation.
\newblock {\em IEEE Photonics Technology Letters}, 29(21):1792--1795, 2017.

\bibitem{liu2016coherent}
L.~Liu, T.~Cheng, K.~Nagasaka, H.~Tong, G.~Qin, T.~Suzuki, and Y.~Ohishi.
\newblock Coherent mid-infrared supercontinuum generation in all-solid chalcogenide microstructured fibers with all-normal dispersion.
\newblock {\em Opt. Lett.}, 41(2):392--395, Jan 2016.

\end{thebibliography}

\end{document}